\documentclass{aa} 
\usepackage{graphicx, amsmath}
\usepackage{verbatim}
\usepackage{txfonts, ulem}
\usepackage{natbib}
\usepackage{hyperref}
\bibpunct{(}{)}{;}{a}{}{,}

\begin{document}

 \title{r-Java 2.0: the astrophysics}

   \author{M. Kostka\thanks{email:mkostka@ucalgary.ca}
          \inst{1}
		\and
          N. Koning
          \inst{1}          
          \and
		  Z. Shand 
          \inst{1}  
          \and        
          R. Ouyed 
          \inst{1}
          \and
          P. Jaikumar
          \inst{2}
          }

   \institute{Department of Physics \& Astronomy, University of Calgary, 
2500 University Drive NW, Calgary, Alberta, T2N 1N4 Canada             
		\and Department of Physics \& Astronomy, California State University Long Beach, 1250 Bellflower Blvd., Long Beach, California 90840 USA         
}

   \date{Received ; accepted }

  \abstract
   {This article is the second in a two part series introducing r-Java 2.0, a nucleosynthesis code for open use that performs r-process calculations and provides a suite of other analysis tools.}
   {The first paper discussed the nuclear physics inherent to r-Java 2.0 and in this article the astrophysics incorporated into the software will be detailed. }
   {R-Java 2.0 allows the user to specify the density and temperature evolution for an r-process simulation.   Defining how the physical parameters (temperature and density) evolve can effectively simulate the astrophysical conditions for the r-process.  Within r-Java 2.0 the user has the option to select astrophysical environments which have unique sets of input parameters available for the user to adjust.  In this work we study three proposed r-process sites; neutrino-driven winds around a proto-neutron star, ejecta from a neutron star merger and ejecta from a quark nova.  The underlying physics that define the temperature and density evolution for each site is described in this work.}
   {In this paper a survey of the available parameters for each astrophysical site is undertaken and the effect on final r-process abundance is compared.  The resulting abundances for each site are also compared to solar observations both independently and in concert.  R-Java 2.0 is available for download from the website of the Quark-Nova Project: \url{quarknova.ucalgary.ca} }
   {}


   \maketitle
   
   \authorrunning{Kostka et al.}
	\titlerunning{Astrophysical r-process sites}

\section{Introduction}
\label{intro}

The rapid neutron capture process (r-process) is believed to be the
mechanism for the nucleosynthesis of about half of the stable nuclei
heavier than iron \citep{Burbidge,Cameron}.  Explosive and neutron-rich astrophysical environments present ideal conditions for the r-process to take place. Possible candidate sites discussed in the literature include the neutrino-driven neutron-rich wind from proto-neutron stars \citep{qian96,Qian03b}, prompt explosions of collapsed stellar cores, \citep{Sumiyoshi,Wanajo,sar12}, neutron star decompression \citep{meyer97,Goriely}, tidal disruption in binary merger events \citep{FRT}, outflows in gamma-ray bursts \citep{Surman}, the LEPP process in low metalicity stars \citep{trav04}, supernova fallback \citep{fryer06}, etc.  Most importantly, abundance data on r-process elements in  metal-poor stars \citep{Sneden03,Truran} and certain radio-nuclides in meteorites \citep{QW08} point toward the distinct possibility of multiple r-process sites.  

An important paper by \cite{beth85} on supernova neutrinos brought the high-entropy environment in neutrino-driven winds from Type II Supernovae (SNe) to the forefront of the discussion about astrophysical sites for the r-process.  Since then, much progress has been made in the modelling of type II SNe and neutrino winds of nascent neutron stars \citep{Woosley,Takahashi,qian96,Cardall,Otsuki,Wanajo01,TBM}. 
The most natural explanation in the neutrino-driven wind scenario is that the observed r-process pattern follows from a superposition of neutron capture events with differing neutron-to-seed ratios and exposure time-scales. A particular challenge for high-entropy winds as an r-process site is that producing the third peak requires extreme values of entropy and dynamic time-scale that are not supported by current hydrodynamic models of Type II SNe explosions \citep[e.g.][]{arc07,QW08,fis10, arc11}. 

Neutron star mergers can provide a much larger neutron-to-seed ratio than type II SNe, which makes for an appealing r-process site.  Recent relativistic hydrodynamic simulations of neutron star mergers have shown that a significant amount of r-process enriched matter can be ejected \citep{janka99,ross04,oec07,gori11}, implying that the solar abundance of r-process nuclei may be influenced by neutron star mergers.  However the neutron star (NS) coalescences time scale which likely ranges from 1 - 1000 Myr$^{-1}$ per Milky Way Equivalent Galaxy \citep{kalo,kaloErr} puts the neutron star merger explanation at odds with enrichment of r-process elements relative to iron observed in metal-poor stars \citep{qian00, arg04}.  

A possible new site for the r-process that is conjectured to be effective in producing the heavy elements beyond the second peak is the quark nova \citep{ODD}. If this scenario occurs in nature, it presents a new possibility for explaining the origin of the heavy elements. \cite{meyer89} and \cite{Goriely} have studied the r-process in decompressing {\it cold} neutron star matter \citep{latt77}, although no specific mechanism for decompression was proposed. In previous work \citep{jaik07,nieb10}, it was suggested that the dynamics of a quark-hadron phase transition inside a neutron star could be sufficiently strong to power the decompression and subsequent ejection of the neutron-rich crust, this is essentially the quark nova scenario.

Clearly there is further study needed to better understand r-process in the Universe and to this end we present r-Java 2.0 an r-process code that is transparent and freely available\footnote{Software and user manual can be downloaded at quarknova.ucalgary.ca} to the nuclear physics and astrophysics community.  This article is meant to display the astrophysics incorporated in r-Java 2.0 and should be considered in conjunction with \cite{kostka} which covered the nuclear physics inherent to the software.  This paper is organized as follows; an overview of r-Java 2.0 is presented in section \ref{rJavaNuc}, the high-entropy wind (HEW) module is described in section \ref{hewSect}, the neutron star merger (NSM) module is laid out in section \ref{nsmSect} and the quark nova (QN) module is discussed in section \ref{qnSect}.  Within each astrophysical module section simulation results are presented.  In section \ref{discuss} the results from each astrophysical scenario is compared to the solar observations.  Finally a summary is presented in section \ref{conc}.

\section{Overview of r-Java 2.0}
\label{rJavaNuc}

The r-process nucleosynthesis code r-Java 2.0 was used in the analysis presented in this paper.  This section will briefly cover the nuclear physics incorporated in the code, for a detailed discussion see \cite{kostka}.  A fundamental feature of r-Java 2.0 is the flexibility afforded to the user to change any nuclear parameter used by the code (i.e. mass model, neutron-capture cross-sections, etc.).  At installation there is default nuclear data provided with r-Java 2.0 and the analysis presented here uses that data set.  The default mass model for r-Java 2.0 is Hartree-Fock-Bogolyubov 21 (HFB21) \citep{hfb21}, with which we utilize the temperature dependant neutron capture cross-sections and photo-dissociation rates as calculated by the TALYS code \citep{TALYS}.  The $\beta$-decay rate and $\beta$-delayed neutron emission of up to three neutrons comes from the work of \cite{moller95}.  The $\alpha$-decay rates are empirically calculated based on binding energy and alpha particle kinetic energy \citep{lang80}.  Spontaneous, neutron-induced and $\beta$-delayed fission are considered in r-Java 2.0 with fission fragmentation explicitly calculated for each parent nuclei.  The spontaneous fission rates are calculated following the methodology laid out by \cite{kod75}.  The $\beta$-delayed fission probabilities were taken from \cite{panov05} and the temperature dependant neutron-induced fission rates \citep{TALYS} provided as defaults in r-Java 2.0 are based on the HFB14 mass model \citep{hfb14}.
  
The temperature range over which r-Java 2.0 is capable of simulating r-process nucleosynthesis is determined by the range of temperature dependant neutron-capture cross-sections (and the corresponding photo-dissociation rates) input by the user.  With the default cross-sections provided with r-Java 2.0 the temperature range is 0.001 $< T_9 < $ 10, where $T_9$ is in units of $10^9 $K.  

Temperature evolution during a simulation is computed through change in entropy per nucleon (s in  $k_{\rm B}$ = 1 units), where entropy is given by
\begin{equation}
\label{Entropy}
s = \frac{11 \pi^2}{45}\frac{T^3}{\rho}+ \frac{\mu_e^2T}{3\rho}+\frac{5}{2m_N}+\frac{1}{m_N}{\rm ln}\left(\frac{g_0(2\pi m_Nk_BT)^{3/2}}{n_nh^3}\right)\ ,
\end{equation}
\noindent where $\mu_e$=$(3\pi^2Y_e\,\rho)^{1/3}$ is the chemical potential of the fully degenerate relativistic electrons, $m_N$=939.1 MeV is the neutron mass and $g_0$=2 (spins) is the statistical weight of the neutron.  The heat generated (or absorbed) by nuclear transmutations is calculated using the same methodology as \cite{hix06} and is added to Eqn. \ref{Entropy} as $\delta s = \delta q/T$.  Finally the whole expression is inverted to get the new temperature at the end of each time step.  As will be discussed in detail in section \ref{hewEjc} under the assumption of radiation dominated entropy the temperature evolution can be simplified to that of adiabatic expansion.  When a user of r-Java 2.0 chooses the HEW module the temperature evolution follows the radiation dominated entropy assumption.  

The initial composition of the system prior to an r-process simulation is free for the user to choose.  Built into the graphical user interface (GUI) of r-Java 2.0 is a table that contains all the nuclear data used by the code including initial mass fractions ($X_{\rm i}$ where $\Sigma_{\rm i=0}^{\rm i = N} X_{\rm i} = 1$ and $N$ is the total number of nuclei in the network).  If the initial composition contains only a few nuclei the user can simply enter the initial mass fraction into the nuclear data table.  However there is as well the option to import the initial mass fractions from a text file, if a large number of nuclei are initially present.  A third option for users of r-Java 2.0 is to first run the nuclear statistical equilibrium (NSE) module and then set the output from the NSE run as the initial abundances for an r-process simulations.  In all cases, prior to (and during) an r-process simulation r-Java 2.0 ensures that conservation of baryon number and charge are upheld.

The density evolution of the r-process site can be freely modified by the user of r-Java 2.0, with the default density evolution profile\footnote{$\rho (t) = \frac{\rho_0}{\left(1+t/2\tau\right)^2}$} coming from the studies of non-relativistic decompression by \cite{meyer97}.  The ability to follow the so-called \textit{classical approach} where density remains constant throughout the r-process simulation is simple to do using r-Java 2.0.  Density evolution and with it the temperature evolution are important features of an astrophysical r-process site.  Thus while the option to define a custom evolutionary scheme is given to the user of r-Java 2.0, we have also incorporated site specific density (and temperature) evolution profiles.  This work will focus on detailing these astrophysical sites which are built into r-Java 2.0 as distinct modules.  The astrophysical sites are: high entropy wind around a proto-neutron stars \citep[example studies;][]{woos92,qianWoos96,thomp01,farouqi10}, ejecta from neutron star mergers \citep[][and others]{freid99b,gori11} and ejecta from quark novae \citep{jaik07,OPJ}.  The details of the calculations for the evolution of physical parameters (in particular temperature and density) for each individual site will be discussed in detail in this work.

\section{High Entropy Winds}
\label{hewSect}

\subsection{Merits as a r-process site}
\label{hewMerits}
The high entropy bubbles in parts of the neutrino-driven winds  surrounding a proto-neutron star make for an intriguing environment for possible r-process nucleosynthesis \citep{Woosley,Takahashi,qian96,Cardall,Otsuki,Wanajo01,TBM}.  \cite{qian08} found that the mass fraction of r-process elements in the solar system was consistent with the amount of material ejected by neutrino-driven winds around a proto-neutron star.  Although \cite{arc11} and \cite{hudepohl} found that based on the latest hydrodynamic simulations the HEW scenario is unable to synthesize heavy r-process elements without artificially increasing the entropy.          

\subsection{Ejection dynamics}
\label{hewEjc}
The r-process nucleosynthesis in the HEW scenario is typically thought to begin at the termination of charged particle ($\alpha$-process) nucleosynthesis \citep{kratz08,far09,farouqi10}.  At charged particle freeze-out, regions of expanding material are considered to have different entropies which can be a measure of the neutron-to-seed ratio \citep{kratz08}.  In this scenario the HEW bubble expands adiabatically, thus temperature is governed by  eqn. \ref{adiaT};

\begin{equation}
\label{adiaT}
T_9 \left( t \right) = T_9 \left( t=0 \right)\frac{R_0}{R_0 + v_{\rm exp}\,t}\,,
\end{equation}

\noindent $R_0$ is the initial radius of the bubble, $v_{\rm exp}$ is the expansion velocity and $t$ is time.  At high temperatures and low matter densities the pressure per unit volume is predominantly due to relativistic particles, such as photons, and the entropy can be considered to be radiation dominated.  \cite{witti} developed an approximation for entropy in this regime;

\begin{equation}
\label{hewEntro}
S = S_{\gamma}\left( 1 + \frac{7}{4}f\left(T_9\right)  \right)
\end{equation}

\noindent where the photon contribution to entropy ($S_{\gamma}$) is

\begin{equation}
\label{phoEntro}
S_{\gamma} = 1.21 \frac{T_9^3}{\rho_5}\,,
\end{equation}
\noindent per baryon in units of $k_{\rm B}$ and $f(T_9)$ is a fit function of the form,

\begin{equation}
\label{ft}
f\left(T_9\right) = \frac{T_9^2}{T_9^2 + 5.3}\,.
\end{equation}

\noindent From eqn. \ref{hewEntro} density ($\rho_5$) can be expressed in units of $10^5$ g cm$^{-3}$ as follows;

\begin{equation}
\label{HEWrho}
\rho_5 \left( t \right) = 1.21 \frac{T_9^3}{S}\left(1+\frac{7}{4}\frac{T_9^2}{\left(T_9^2 + 5.3\right)}\right)\, .
\end{equation}

\noindent Within the HEW module of r-Java 2.0 $T_0$, $S$, $R_0$ and $v_{\rm exp}$ are free parameters for the user to adjust.  The temperature and density evolution described in this section is used by r-Java 2.0 when the HEW module is selected by the user.

\subsection{Simulation Results}
\label{hewRes}
In this work we consider the final r-process nuclei abundances for the HEW scenario for a range of physical parameters given different initial abundance distributions.   

\subsubsection{Varying Initial Abundances}
\label{hewAbdRes}
\begin{figure}
\includegraphics[scale = 0.75]{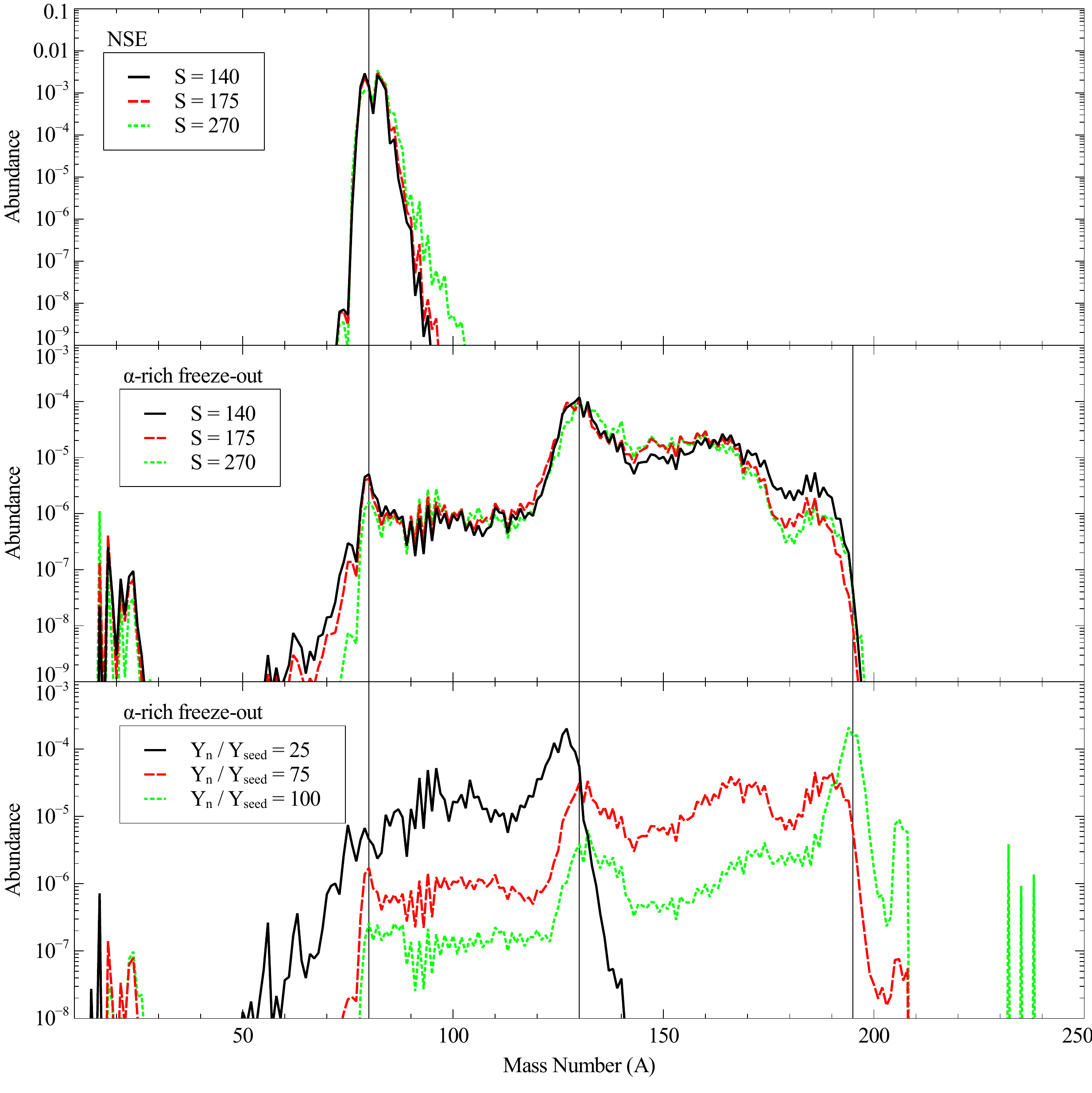}
\caption{\textbf{High-entropy wind.}  See section \ref{hewRes} for details. \textit{Top:}  The black solid line denotes an entropy (S) of 140 and expansion velocity ($v_{\rm exp}$) of 3$\times10^4$ km s$^{-1}$.  The red dashed line denotes S = 175 and $v_{\rm exp}$ = 1.5$\times10^4$ km s$^{-1}$.  The green dotted line denotes S = 270 and $v_{\rm exp}$ = 3.8$\times10^3$ km s$^{-1}$.  NSE determined initial abundances.  \textit{Middle:} Same as top panel with $\alpha$-rich freeze-out initial abundances.  \textit{Bottom:}  The black solid line denotes a neutron-to-seed ratio of 25, the red dashed line a neutron-to-seed ratio of 75 and the green dotted line a neutron-to-seed ratio of 100.  Each simulation started from $\alpha$-rich freeze-out abundances.}
\label{HEWcomp}
\end{figure}

The cases seen in the top panel of Fig. \ref{HEWcomp} considers r-process in the HEW scenario starting from a nuclei distribution as determined by NSE.  Each result plotted in the top panel of Fig. \ref{HEWcomp} began with an initial temperature of $3\times10^9$ K and an initial electron fraction of 0.35.  The expansion velocity and entropy pairs, $(v_{\rm exp}:S) \in \lbrace$ (3750 km s$^{-1}$:270), (1.5$\times10^4$ km s$^{-1}$:175) and (3.0$\times10^4$ km s$^{-1}$:140) $\rbrace$ were taken from Fig. 4 of \cite{farouqi10}.  The chosen entropy and initial temperature were used in eqn. \ref{HEWrho} to determine the initial density.  Based on density, temperature and electron fraction the initial abundance was calculated assuming NSE with the inclusion of coulomb interactions.  For each case studied in the top panel of Fig. \ref{HEWcomp} the neutron-to-seed ratio, as determined by NSE, was approximately five and the initial nuclei abundance distribution was described by a peak ranging from $75 \lesssim$A$\lesssim 80$.  For each entropy and expansion velocity pair the r-process was capable of shifting the distribution to the heavy-side of the A = 80 magic number with the highest entropy case (S = 270) having the longest heavy-side tail which extended slightly past A = 100.   

The study of HEWs as an r-process site has been shown to be dependent on the nuclei abundance at the termination of the $\alpha$-process \citep{woos92, kratz08, farouqi10}.  In the case of \textquotedblleft$\alpha$-rich freeze-out" not all of the $\alpha$-particles are consumed during the $\alpha$-process and a significant abundance of heavy seed nuclei are created which can subsequently capture neutrons in the r-process.  

Using the data provided in \cite{farouqi10} we are able to reconstruct an initial r-process abundance distribution that approximates an $\alpha$-rich freeze-out.  With the assertion that at the termination of the $\alpha$-process the temperature of the HEW bubble is 3$\times10^9$K and the neutron density is $\sim10^{27}$ cm$^{-3}$ we can assume that abundance along isotopic chains will follow the waiting point approximation.  Using the work of \cite{qian03} the most abundant nucleus in an isotopic chain is that whose neutron separation is closest to

\begin{equation}
\label{SnApprox}
\overline{S}_{\rm N} = T_9 \left( 2.79+0.189\left(\log\left(\frac{10^{20} cm^{-3}}{n_n}\right)+\frac{3}{2}\log\left(T_9\right)\right)\right)\, {\rm MeV.}
\end{equation}

\noindent We then set the abundance for each mass number as described in Fig. 4 of \cite{farouqi10} to the nucleus which best approximated eqn. \ref{SnApprox}.  With this $\alpha$-rich freeze-out initial abundance distribution we ran two comparisons; one considering the effect of changing entropy and expansion velocity pairs while holding the neutron-to-seed ratio fixed and the other exploring a range in neutron-to-seed ratios.

The middle panel of Fig. \ref{HEWcomp} displays simulation results using the same three entropy and expansion velocity pairs used the top panel; however for the simulation results shown in the middle panel the initial nuclei abundance distribution was found using our $\alpha$-rich freeze-out approximation.  For each simulation seen in the middle panel of Fig. \ref{HEWcomp} the neutron-to-seed ratio was held fixed (Y$_{\rm n}/$Y$_{\rm seed}$ = 50) along with the initial temperature ($T_0 = 3\times 10^9$K) and initial electron fraction ($Y_{\rm e}$ = 0.45).  Each simulation result shown in the middle panel of Fig. \ref{HEWcomp} displays roughly the same distribution, a strong A = 130 peak, a weak A = 80 peak and a cluster of nuclei around A = 25 which are the result of neutron capture onto $\alpha$-capture products.  The most noticeable difference in r-process abundance yields between simulations shown in the middle panel of Fig. \ref{HEWcomp} is that the lowest entropy case (S = 140) investigated provided the largest abundance of heavy nuclei.  This result is due to the fact that as eqn. \ref{HEWrho} dictates, a lower entropy implies a higher initial density (for the same initial temperature) and thus in the S = 140 case the neutron density remained high enough for successful neutron capture longer than the other two cases studied (S = 175 and S = 270).  This result seems counter-intuitive as conventionally higher entropy implies heavier r-process abundance yields.  However the convention is that higher entropy as well implies a larger neutron-to-seed ratio which was not the case for this study in which we chose to fix the neutron-to-seed ratio for each simulation displayed in the middle panel of Fig. \ref{HEWcomp}.  Similar results of lower entropy r-process giving heavier elements have been seen by \cite{qian08}.

\subsubsection{Varying Physical Parameters}
\label{hewPhysRes}
The effect of varying the neutron-to-seed ratio within the HEW scenario can be seen in the bottom panel of Fig. \ref{HEWcomp}.  For these simulations   the initial temperature was set to 3$\times10^9$ K, the initial electron fraction was 0.45 and the expansion velocity was 1.5$\times10^4$ km s$^{-1}$.  The initial nuclei distribution was determined following the $\alpha$-rich freeze-out methodology discussed earlier, however the seed abundances were scaled differently in order to achieve three different neutron-to-seed  ratios; Y$_{\rm n}$/Y$_{\rm seed}$ = 25 (S = 175), 75 (S = 250) and 100 (S = 270).  For each neutron-to-seed ratio an entropy was chosen to be in accordance with table 5 of \cite{farouqi10}.  The final nuclei abundance distribution for the lowest neutron-to-seed ratio (Y$_{\rm n}$/Y$_{\rm seed}$ = 25) plotted in the bottom panel of Fig. \ref{HEWcomp} displays a peak at approximately A = 126.   For this neutron-to-seed ratio the r-process is capable of reaching the A = 130 waiting point, but not strong enough the significantly produce heavier nuclei.  For the Y$_{\rm n}$/Y$_{\rm seed}$ = 75 case, the nuclei in the final distribution are concentrated in three peaks of roughly the same height at A = 130, 180 and 190.  The heaviest nuclei formed in the Y$_{\rm n}$/Y$_{\rm seed}$ = 75 case are nuclei which have built up on the light-side of the A = 195 magic number, with only a small fraction of heavier nuclei formed.  As shown in the bottom panel of Fig. \ref{HEWcomp} with Y$_{\rm n}$/Y$_{\rm seed}$ = 100 the r-process in the HEW scenario is capable of forming transuranic elements as well as a strong A = 195 peak.  

R-process in the HEWs around a proto-neutron star is a delicate subject requiring exhaustive study of when the environment is favourable and how the seed nuclei are formed \citep{thie11,arc12}.  For this reason we will leave the undertaking of a thorough examination of the radiation dominated approximation of entropy as well as other aspects of HEWs in the context of r-process for a future paper.

\section{Neutron Star Mergers}
\label{nsmSect}

\subsection{Merits as r-process site}
\label{nsmMerits}

Hydrodynamic simulations of NSM showing that an appreciable amount of matter could become gravitationally unbound \citep[e.g.][]{janka99,ross04,oec07,gori11} piqued interest in NSM as a possible r-process site.  \cite{gori11} used relativistic NSM models to find that $\sim 10^{-3}$M$_{\odot}$ could be ejected in both symmetric and asymmetric systems.  This amount of ejected matter coupled with the expected rate within our Galaxy of $10^{-5}$yr$^{-1}$ \citep{phin91,arz99,blec02} implies that NSM events may play a significant role in the formation of galactic r-process elements.

\subsection{Ejection Dynamics}
\label{nsmEjc}

Motivated by the results from hydrodynamic simulations, we built into r-Java 2.0 a tool for the study of r-process in the context of NSMs.  While common practice is to utilize the results of hydrodynamic simulations to define the density evolution of the material undergoing r-process nucleosynthesis, we felt that an analytic approach could allow users to study a wide array of NSM scenarios without being reliant on obtaining the results from hydrodynamic simulations.  
For the NSM module of r-Java 2.0 the temperature evolution is found through change in entropy and follows the prescription described in Sect. \ref{rJavaNuc}.  The density evolution used in the NSM module assumes the pressure-driven expansion of ejected NS matter and considers the stiffness of the equation of state in determining the dynamic evolution of the ejecta.   

The NSM environment in r-Java 2.0 assumes the equation of state of the NS material to be polytropic, where pressure is related to density as $P = K\rho^{\gamma}$ with $\gamma = \frac{n+1}{n}$, with the polytropic index ($n$) a free parameter and $K$ determined from initial conditions.  The density evolution of the chunk of ejected neutron star matter follows $\rho(t) = \rho_0(R_0/R(t))^3$, where the initial density ($\rho_0$) and initial radius ($R_0$) of the chunk are free input parameters.  The time-evolution of the radius, $R(t)$, of the chunk of matter is found by numerically integrating the Newtonian equation of motion.  

\begin{equation}
\label{NSMeom}
\frac{{\rm d}^2R}{{\rm d}t^2} = -\frac{G\,M_{\rm c}}{R^2} - \rho^{-1}\frac{{\rm d}P}{{\rm d}R} + G\,M_{\rm ns}\left(\frac{1}{r_{\rm 
c}^2} - \frac{1}{\left(r_{\rm c} + R \right)^2}\right).
\end{equation}

\noindent Where $G$ is Newton's gravitational constant, $P$ is the internal pressure of the chunk, $\rho$ is the average density, $M$ denotes mass with the subscript \textit{c} denoting that of the chunk of ejected material and \textit{ns} that of the neutron star and $r_{\rm c}$ represents the radial position of the center of the ejected chunk.  The first term in eqn. \ref{NSMeom} represents the inward acceleration due to the self-gravity of the chunk.  The self-gravity term can be expressed in terms of density by considering $M_{\rm c} = 4 \pi \rho R^3/3$ leaving the acceleration due to self-gravity as;

\begin{equation}
\label{Asg}
a_{\rm sg} = -\frac{4\,\pi}{3}G\,\rho\,R .
\end{equation}

\noindent The second term in eqn. \ref{NSMeom} defines the pressure driven acceleration.  Since we assume a polytropic equation of state the acceleration can be found analytically as;

\begin{equation}
\label{Ap}
\begin{aligned}
a_{\rm p} & = & -\rho^{-1}\frac{\partial P}{\partial \rho}\frac{{\rm d \rho}}{{\rm d} R} \\
& = & - K \gamma \rho^{\gamma -2} \frac{d}{d R}\left(\rho _0\left(\frac{R_0}{R}\right)^3\right)\\
& = & \frac{3 \gamma P}{\rho R}.
\end{aligned}
\end{equation}

\noindent The third term in eqn. \ref{NSMeom} represents the effect of tidal stretching, which under the assumption that $r_{\rm c} \gg R$  can reduce to $R\left(\tau_{\rm esc} + \frac{3}{2}t\right)^{-2}  $ where $\tau_{\rm esc}$ is the expansion timescale.  Using an analytic expression for the equation of state of the ejected NS matter has the advantage of allowing us to incorporate the effect of stiffness into the density evolution. By considering the elasticity we can incorporate a first order approximation to the material's resistance to tidal stretching.  This is done using the bulk modulus which can be expressed as $B = \rho \left({\rm d}P/{\rm d}\rho \right)$, in units of pressure.  Multiplying the bulk modulus by the surface area of the ejected matter and dividing by the mass yields an acceleration that can be incorporated into eqn. \ref{NSMeom}.  This is a resistive acceleration that is stronger with stiffer equations of state.  

\begin{figure}
\includegraphics{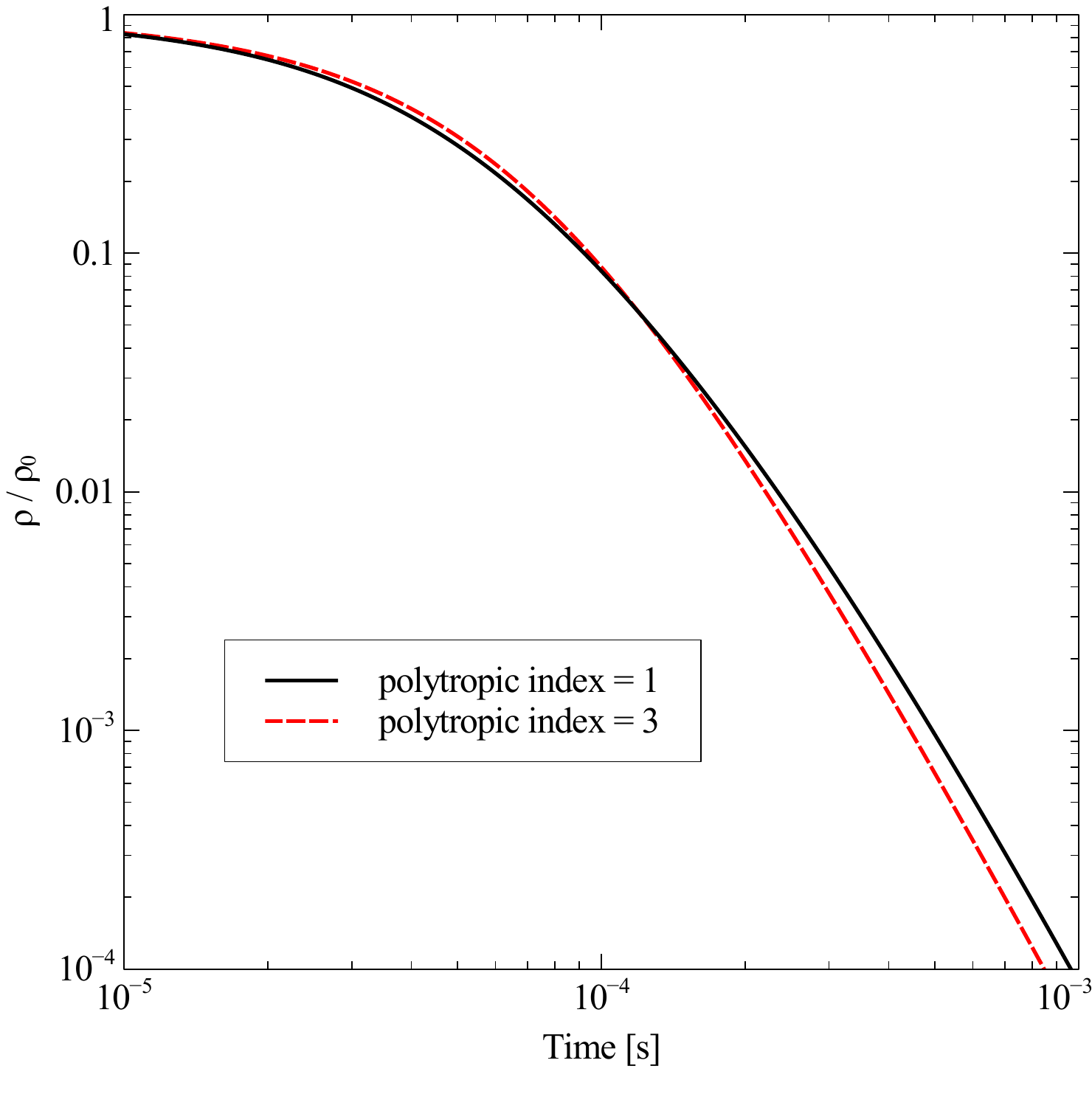}
\caption{Using the neutron star merger module the density evolution of ejecta with two different polytropic indices (n) are compared.  The black solid line denotes relativistically degenerate matter (n = 1) and the red dashed line represents non-relativistically degenerate matter (n = 3). }
\label{NSMrho}
\end{figure}

The density evolution of NSM scenarios defined by different polytropic indices ( n = 1 and n = 3) can be seen in Fig. \ref{NSMrho}.  The density of the relativistically degenerate system (polytropic index of 1) drops faster initially, as the pressure-gradient acceleration is stronger for this case.  However as tidal stretching becomes the dominant acceleration term in eqn. \ref{NSMeom} the density of the softer, non-relativistically degenerate (n = 3) system begins to drop more quickly.  The crossing point between the two density evolutions seen in Fig. \ref{NSMrho} is roughly the expansion timescale, which for this comparison was $2\times10^{-4}$ seconds.

\subsection{Simulation Results}
\label{NSMresults}

Using the NSM module of r-Java 2.0 we investigated the effect on final r-process abundances of changing both the physical parameters that define the NSM and the initial nuclei abundance distribution.  Varying initial temperature, expansion velocity and the initial density are studied.  As well the final nuclei distributions for different polytropic indices are shown.  The three different initial abundance scenarios investigated are; nuclei in ejecta are initially in NSE, the presence of an iron group seed and initial dissociation into neutrons, protons and alpha particles.  

\subsubsection{Varying Physical Parameters}
\label{NSMphysRes}

\begin{figure}
\includegraphics[scale = 0.75]{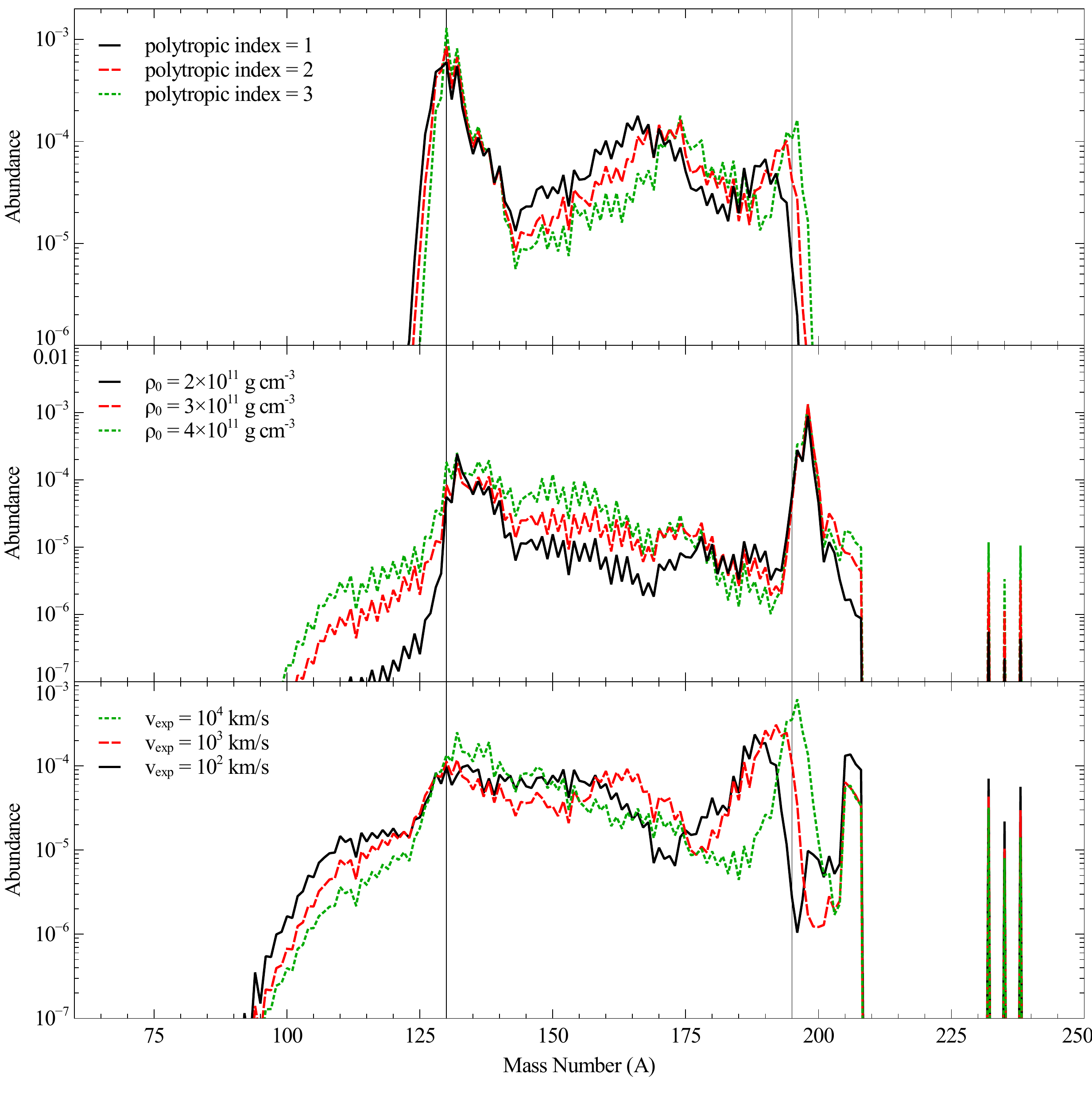}
\caption{\textbf{Neutron star merger.}  Final r-process abundances are compared for varying physical parameters.  See section \ref{NSMphysRes} for details of simulations.  \textit{Top:} Varying polytropic indexes (n); n = 1 (black solid line), n = 2 (red dashed line) and n = 3 (green dotted line).  \textit{Middle:}  Varying initial densities ($\rho_0$); $\rho_0 =  2\times10^{11}$g cm$^{-3}$ denoted by the black solid line, $\rho_0 =  3\times10^{11}$g cm$^{-3}$ by the red dashed line and $\rho_0 =  4\times10^{11}$g cm$^{-3}$ by the green dotted line.  \textit{Bottom:}  Varying expansion velocity ($v_{\rm exp}$); $v_{\rm exp} = 10^2$km s$^{-1}$ denoted by the black solid line, $v_{\rm exp} = 10^3$km s$^{-1}$ by the red dashed line and $v_{\rm exp} = 10^4$km s$^{-1}$ by the green dotted line   }
\label{NSMcompPhys}
\end{figure}

The effect on r-process abundance of varying the polytropic index can be seen in the top panel of Fig. \ref{NSMcompPhys}.  The simulation results shown in the top panel of Fig. \ref{NSMcompPhys} each start from the same initial conditions with  $Y_{\rm e,0}$ =0.25.  A trend of heavier nuclei produced for softer polytropic indices can be seen for the initial conditions chosen for this comparison.  Fro each of these simulation the r-process is not capable of pushing through the A = 195 waiting point for any of the chosen polytropic indices.  We found that the influence of polytropic index on nuclei abundances is overwhelmed by fission recycling when the initial electron fraction is low ($Y_{\rm e,0} \sim 0.1$).  For the case of low initial electron fraction, numerous fission recycles cause the initial neutron-to-seed to be a much more dominant attribute in determining final nuclei abundance distribution.  In an effort to be consistent with the r-process in NSM work done using hydrodynamic simulations \citep[eg.][]{RufHydro, gori11}, for the remainder of this study of the NSM module of r-Java 2.0 we will limit ourselves to a polytropic index of one and low electron fractions.

For each of the results plotted in the middle panel of Fig. \ref{NSMcompPhys} the same initial physical conditions were used, with the exception of mass density which was varied.  The initial abundances for each run was determined using NSE.  For the highest density ($\rho_0 = 4\times10^{11}$g cm$^{-3}$) simulation the r-process slows significantly as the neutron-to-r-process products ratio ($Y_n/Y_{r}$) approaches the minimum threshold in r-Java 2.0 which is one.  Computationally this is seen as r-Java 2.0 being able to take very large time-steps and still satisfy its precision requirements of dY(Z,A)$_{\rm max} < 10\%$ per time-step.  For the $\rho_0 = 3\times10^{11}$g cm$^{-3}$ simulation result plotted in the middle panel of Fig. \ref{NSMcompPhys} the temperature and neutron density drop to a level that the r-process is no longer efficient and the simulation \textit{artificially} stops at the user input simulation duration.  For the lowest density case studied ($\rho_0 = 2\times10^{11}$g cm$^{-3}$)  the temperature drops below the minimum threshold (as defined by the temperature dependant rates and cross-sections used in this work) while the neutron density was still high.  For each of the density cases plotted in the middle panel of panel of Fig. \ref{NSMcompPhys} the final nuclei abundance displays a strong peak at A = 195 and a peak at A = 130.  The abundance of the intermediate region between these two peaks generally increases with increasing density.   

As the expansion velocity (v$_{\rm exp}$) along with the initial radius define the expansion timescale, it is expected that these variables should have a strong impact on r-process abundance yields in the context of NSM \citep{gori11}.  The bottom panel of Fig. \ref{NSMcompPhys} highlights the importance of v$_{\rm exp}$ on r-process abundance yields as each simulation is subject to the same initial abundances (as determined by NSE) and initial physical conditions, save v$_{\rm exp}$, and each case studied shows significantly different final abundance distributions.  The most rapidly expanding scenario (v$_{\rm exp} = 10^4$ km s$^{-1}$) displays the largest peak at A = 195.  The relative heights of the A $\sim$ 195 peak can be thought of as inversely proportional to the amount of fission recycling having taken place during the simulation.  For the slower expanding NSM ejecta the physical environment remains favourable for the r-process for longer allowing more material to be pushed into the fissionable regime, thus the v$_{\rm exp} = 10^2$ km s$^{-1}$ case displays the lowest A$\sim$195 peak and the largest abundance of $^{232}$Th, $^{235}$U and $^{238}$U.  The A $\sim$ 130 peak is the strongest in the v$_{\rm exp} = 10^4$ km s$^{-1}$ case, not due to fission recycling, but rather due to material being still \textit{caught} by the A = 130 waiting point.

\subsubsection{Varying Initial Abundances}
\label{NSMiniRes}
 
When comparing the effect of initial abundances on final r-process abundance yields within the NSM scenario, each simulation started with the same physical parameters; T$_0 = 3 \times 10^9$K, $\rho_0 = 8 \times 10^{11}$g cm$^{-3}$, n = 1 and v$_{\rm exp} = 10^4$km s$^{-1}$. Then for each initial abundance scheme (initially in NSE, Fe group seed and dissociation into neutrons, protons and alpha particles) Y$_{\rm e}$ was varied between; 0.1, 0.15 and 0.2.

\begin{figure}
\includegraphics[scale = 0.75]{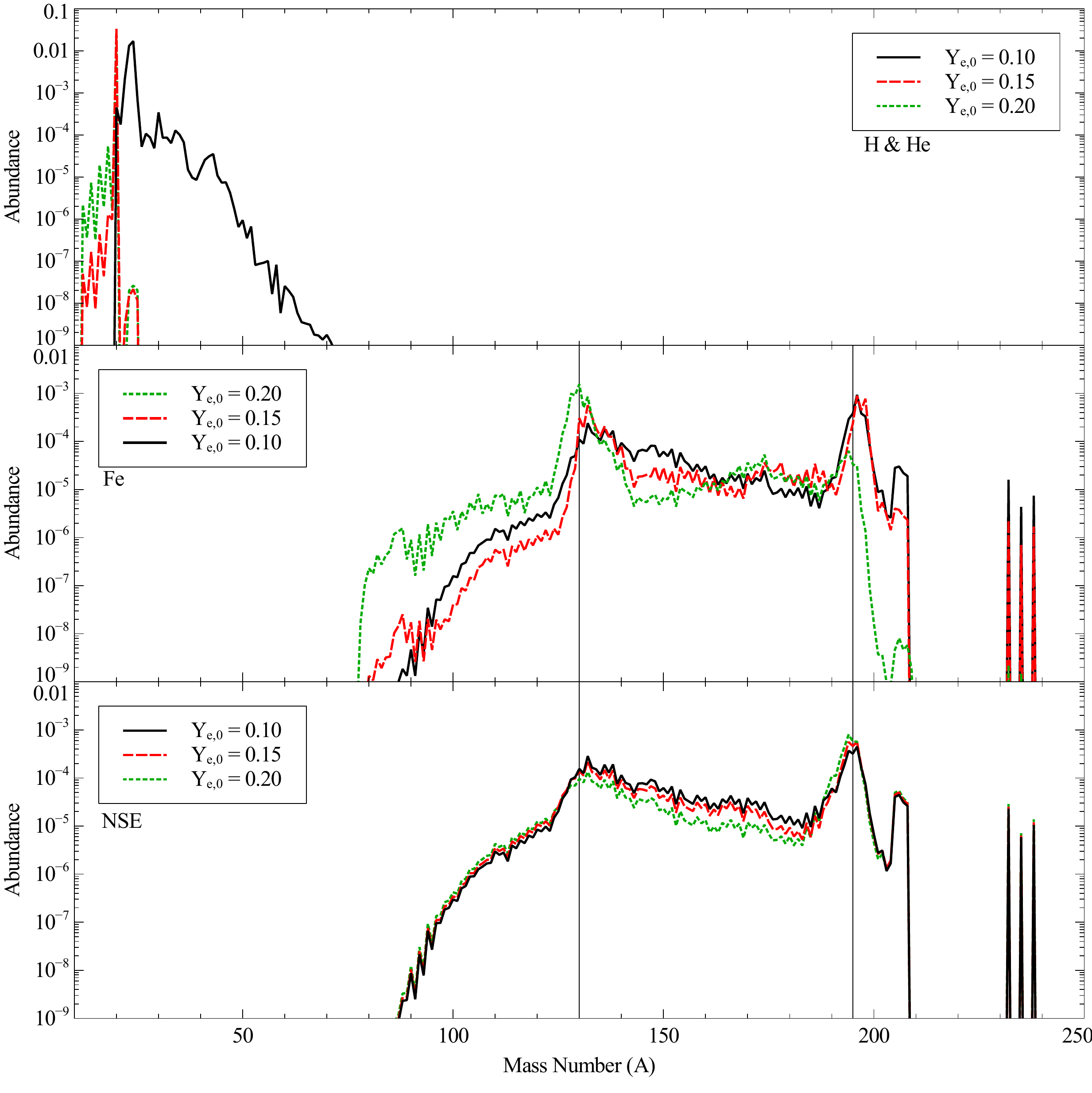}
\caption{\textbf{Neutron star merger.} The final nuclei abundances are plotted for different initial electron fractions ($Y_{\rm e,0} = 0.1$ denoted by the black solid line, $Y_{\rm e,0} = 0.15$ by the red dashed line and $Y_{\rm e,0} = 0.2$ by the green dotted line.  See Sect. \ref{NSMiniRes} for details on initial physical parameters.  \textit{Top:}  Initially only neutron, protons and $\alpha$-particles present.  \textit{Middle:}  An initial seed of iron group isotopes.  \textit{Bottom:} The initial abundances determined using NSE. }
\label{NSMcompInit}
\end{figure}

The case in which initially the NSM ejecta has been dissociated to neutrons, protons and alpha particles is displayed in the top panel of Fig. \ref{NSMcompInit}.  For the Y$_{\rm e,0}$ = 0.15 and 0.2 cases the r-process is unable to get running in a significant manner and the final nuclei distribution is predominantly due to alpha-capture reactions.  As for the  Y$_{\rm e,0}$ = 0.1 simulation, the neutron-to-seed ratio is sufficiently high that after the temperature is too low for significant alpha-capture the r-process can produce nuclei up to A $\sim$ 70.  This work should be revisited with a more robust treatment of alpha-capture, as r-Java 2.0 only considers the triple alpha reaction and alpha-capture onto $^{12}$C and $^{16}$O with the cross-sections determined by \cite{caugh88}.  

The final abundance distribution for the case in which the NSM event has a seed of iron group isotopes present at the beginning of r-process nucleosynthesis can be seen in the middle panel of Fig. \ref{NSMcompInit}.  As with the NSE determined initial abundance case seen in the bottom panel of Fig. \ref{NSMcompInit} fission recycling plays a role, however to less of a degree for the higher Y$_{\rm e,0}$ cases (0.15 and 0.2).  The reduced influence of fission recycling is a product of the average atomic mass of the seed being significantly lower ($\langle$A$\rangle$ $\sim 60$ for the iron group seed and $\langle$A$\rangle$ $\sim$ 100 for NSE).  By the time fission recycling takes hold the neutrons are nearly exhausted for the Y$_{\rm e,0}$ = 0.15 and 0.2 cases.  The final abundance distribution for the  Y$_{\rm e,0}$ = 0.2 simulation is predominantly nuclei in the A = 130 with two small peaks at A $\sim$ 175 and A = 195.  The r-process in the Y$_{\rm e,0}$ = 0.2 simulation run produced only a very small amounts of $^{232}$Th and $^{235}$U.  For the Y$_{\rm e,0}$ = 0.15 simulation seen in the bottom panel of Fig. \ref{NSMcompInit} the final abundances are dominated by peaks at A = 130 and A = 195.  When compared to the results of the Y$_{\rm e,0}$ = 0.1 case the A = 195 peak in the Y$_{\rm e,0}$ = 0.15 case is quite similar in height and breadth, while the A = 130 peak = is narrower.  The increased breadth of the A = 130 peak in the Y$_{\rm e,0}$ = 0.1 simulation is a result of more fission recycling.  

The bottom panel of Fig. \ref{NSMcompInit} displays the NSM scenario in which the initial nuclei abundance distribution is determined by NSE with the aforementioned temperature, density and electron fraction.   The similarity of the final abundances is due to the fact that for each Y$_{\rm e}$ studied, the neutron-to-seed ratio is sufficiently high such that fission recycling becomes dominant and shapes the final abundance distribution.  The slight differences are a product of where the bulk of the material is along the r-process path when $Y_n/Y_{r}$ drops below one.

\section{Quark Novae}
\label{qnSect}

\subsection{Merits as an r-process site}
\label{QNmerits}

The neutron star ejecta is very efficient at producing elements above A $\sim$130 because the seed-nuclei from the ejected crust are already neutron-rich and dynamical time-scales are short ($\tau_{\rm dyn}\sim$0.1-1 ms). This is borne out by Fig.9 and Fig. 10 of \cite{jaik07}, which show that elements above $A\sim$130 are easily produced while elements below $A\sim$130 are deficient.  Ejection events from isolated neutron stars can happen whenever there is a quark-hadron phase transition in newly-born neutron stars formed from the core collapse of (a certain fraction of) the first generation of stars $M\sim$ 25-40$M_{\odot}$ \citep{OPJ}.  Since such stars have typical lifetimes of $10^6$ yrs, the ejection events are likely to be  frequent in the early universe.   This could be frequent enough to explain the observations of early r-process enrichment in metal-poor stars, although this needs to be checked with a detailed chemical evolution study. Further, the QN rate likely decreases towards the present time, if the initial-mass function (IMF) evolves to a less top-heavy one than in the past. Thus, although QN can produce as much as $10^{-2}M_{\odot}$ of r-process material \citep{vogt04,ORV}, such ejection events are not expected to produce large chemical inhomogeneities that would put them in obvious conflict with the observed decrease in scatter of r-process elements over time ($[r]/{\rm Fe}\sim$ 0.2-0.3 dex; \cite{arg04}). 

\subsection{Ejection mechanism}
\label{QNeject}
Our focus of this application of r-Java 2.0 is to consider the Quark-Nova (henceforth QN) as just one of several possible ejection mechanisms for neutron star material. First proposed by \cite{ODD} as a means to power the central engine of gamma-ray bursts, it has since been developed in more detail \citep{KO,KOJ} and discussed in many contexts such as GRBs \citep{OSan, ORV, SNO}, magnetic field decay of AXPs and SGRs \citep{NOL,koning}, collapse to black holes \citep{Manj}, ultra-luminous SNe  \citep{LO,06gy} and reionization from the first stars \citep{OPJ}. The QN converts gravitational energy and nuclear binding energy partly into internal energy (heat) and partly into kinetic energy, with the majority energy release taken by neutrinos. Due to high temperatures and the higher density of quark matter compared to neutron matter, neutrinos are trapped, thus raising the temperature of the adiabatically collapsing quark core to about 10 MeV (approximately $10^{11}$K). In \cite{KOJ}, neutrinos emitted from the conversion of up and down quark matter to
strange matter were assumed to transport the energy into the outer
regions of the star, leading to mass ejection. However,
with neutrino-driven mass ejection, most of the neutrinos that can
escape the core lose their energy to the star's outer layers of
neutron matter in the form of heat. Consequently, mass ejection is
limited to about $10^{-5}M_{\odot}$ for compact quark cores of size
(1-2) km \citep{KOJ}. A more attractive possibility is that of the  
photon and relativistic $e^+e^-$ fireball formed at the base of the hadronic crust due to the underlying hot quark matter \citep{vogt04}, 
which can impart sufficient kinetic energy to the outer layers, 
including the crust of the star (about $10^{-2}M_{\odot}$ can be ejected at maximum efficiency). Mass ejection may also happen through
shock waves propagating outwards in a deflagration, or through a detonative conversion to strange quark matter, as the most recent numerical analysis on this matter seems to suggest \citep{nieb10}. The neutron-to-seed ratio in the ejected crust depends on the initial choice of $Y_e$ and $\tau$, but can easily be $\approx$1000 (Table 2 in Meyer 1989).  The neutron-rich ejecta is heated rapidly by the exploding fireball and subsequent nuclear beta-decay further increases this temperature to $T\geq 4\times 10^9$K, making for ideal r-process conditions in the ejecta.  Spallation is another nucleosynthesis process achievable by the QN that would effect the final nuclei abundances.  If, on the order of days, a QN were to follow its precursory core collapse supernova the inner region of the supernova envelope would likely undergo spallation due to the bombardment of QN ejecta material \citep{amir}.  This process and its resultant chemical signature is an avenue of future development for r-Java 2.0

\subsubsection{Input parameters} 
\label{QNinput}
Usually r-process simulations are run with an initial choice of three input parameters: entropy/nucleon $s$, electron fraction $Y_e$, and expansion time-scale $\tau$. We have created a module in r-Java 2.0 tailored to the QN scenario that can be used in the simplest possible manner by astrophysicists and nuclear physicists alike.  We choose to select three global parameters: ejected mass $M_{\rm QN,ejecta}$, neutron star mass $M_{\rm NS}$ and neutron star radius $R_{\rm NS}$ as input parameters. We show below how our code automatically maps these three inputs to the conventional microscopic inputs and Table \ref{qnTable} displays the calculated inputs for a set of ejecta masses.   

As the density and temperature evolution of the ejecta is different depending on the speed of the ejecta (which depends in turn on the strength of the explosion), we begin with an estimate of the initial Lorentz factor of the shell. The user-specified total ejected mass $M_{\rm QN, ejecta}$ determines the inner radius from which neutron-rich matter is ejected. For this, we invert the equation $M_{\rm QN, ejecta} = \int_{R_{\rm in}}^{R_{\rm NS}} \rho_{\rm NS} 4\pi r^2 dr$
to find  the ejecta inner radius $R_{\rm in}$ and thus the ejected shell's initial thickness $\Delta R_0 = (R_{\rm NS}  - R_{\rm in})$, where typically $R_{NS}\sim 10$km is specified by the user. As discussed in \S \ref{QNeject}, the range of ejecta mass in the QN is
 $10^{-5} M_{\odot} < M_{\rm QN,ejecta} < 10^{-2}M_{\odot}$ which corresponds to a range in $R_{\rm in}$, $9.3\ {\rm km} < R_{\rm in} < 9.7\ {\rm km}$ for a canonical neutron star with $M_{NS}=1.4M_{\odot}$ and $R_{NS}=10$km.  To determine the entropy Eqn.(\ref{Entropy}) is used with,  $\rho_0$= $\rho_{\rm av.}$, $Y_{\rm e, 0}$= $Y_{\rm e, av.}$
where $\rho_{\rm av.}$ = $M_{\rm QN,ejecta}/V_{\rm ejecta., 0}$
is the ejecta average density with $V_{\rm ejecta., 0}$ as the initial volume of the ejecta and $Y_{\rm e, av.}$= $\int_{\rm R_{\rm in}}^{R_{\rm drip}}
Y_{\rm e}(\rho) \rho dV_{\rm ejecta}/ \int_{\rm R_{\rm in}}^{R_{\rm drip}}
 \rho dV_{\rm ejecta}$ is electron
 fraction. The upper integration limit $R_{\rm drip}$,
 in $Y_{\rm e, av.}$  is found by the code from $ \rho (R_{\rm drip})$ = $4\times 10^{11}$ gm
 cm$^{-3}$ and yields $R_{\rm drip}\sim 9.72$ km ($ > R_{\rm in, max}$=9.7 km).

For the interior structure of the neutron star (the QN progenitor), we adopt the following parameterization for the density profile of the neutron star which follows from the work of \cite{latt77}. This profile coresponds to the BPS equation of state \citep{Crust} at low density, matched to the BBP equation of state \citep{Baym} at densities up to nuclear saturation density. This was also used in \cite{jaik07}:

          \begin{equation}
          \log(\rho_{\rm NS}(R)) \simeq a_0 - a_1 R + a_2  R^2 - a_3 R^3\ ,
          \end{equation}

\noindent where $R$ is the neutron star interior radius in kilometers (where $R > R_{\rm drip}$), $\rho_{\rm NS}$ is in g/cc and $a_0= 100.0, a_1= 78.04,a_2= 14.172,a_3= 0.7283$.

In order to determine the Lorentz factor ($\Gamma_{\rm QN}$) first we consider the fraction of total energy of the QN (gravitational and nuclear binding energy) that is transformed into kinetic energy of the ejecta ($E_{\rm QN}^{\rm KE} = \zeta_{\rm  KE} ~ E_{\rm QN}^{\rm tot} $), characterized by $\zeta_{\rm KE}\sim 0.1$.  Then $\Gamma_{\rm QN} = E_{\rm QN}^{\rm KE}~/~\left(M_{\rm QN, ejecta}~c^2\right)$, which is expressed explicitly as

\begin{equation}
\label{lorentz}
\Gamma_{\rm QN} =\zeta_{\rm  KE}\left( \frac{\Delta R_0}{R_{\rm
NS}}\right)\left(\frac{R_{\rm
Schw.,\odot}}{R_{NS}}\right)\left(\frac{M_{NS}}{M_{\odot}}\right)^2\slash\left(\frac{M_{\rm QN,ejecta}}{M_{\odot}}\right)\, .
\end{equation}

We find that if $M_{\rm QN,ejecta}\sim 10^{-3}$M$_{\odot}$ or larger we are in the non-relativistic regime.

For the electron fraction, we are using a parameterized formula for the density dependence of Ye (extracted from Table 1 in \cite{meyer89})

\begin{equation}
Y_{\rm e}(\rho) \simeq b_0 - b_1  (\log\rho) + b_2 (\log\rho)^2 - b_3 ( \log\rho )^3\ ,
\end{equation}
with $\rho$ in g/cc and
$b_0= 155.66, b_1= -35.401,b_2= 2.6856,b_3= -0.067933$.  

Having established the starting entropy/nucleon and initial $Y_e$, we still need to
determine the expansion time-scale. This is related to the initial Lorentz factor (Eqn.(\ref{lorentz}))
and the ejecta shell expansion.  Previous works \citep{jaik07} have used an expression where the
ejecta is driven by internal pressure:
\begin{equation}
\label{radius}
 \frac {d^2R}{dt^2} = \frac{\alpha}{\rho} \frac{P}{R} = \alpha \frac{c_{\rm s}^2}{R} \ .
\end{equation}

where $c_{\rm s}$ is the sound speed. 
In a QN however, the initial expansion is in general much faster than implied by Eqn.(\ref{radius}).  In order to calculate the density evolution, first the thickness of the shell ($\Delta r$) must be determined as a function of time (in the shell's frame $\Delta r$ expands linearly at the sound speed).  Since for dense matter the sound speed is a function of density two regimes must be considered, one for degenerate matter and another for non-degenerate matter, where the transition density is $\rho_{\rm tr} \sim 2\times 10^6$ g/cc.  A derivation of the evolution of the thickness of the ejected shell in its frame can be found in the Appendix of this paper.  It is necessary to consider the shell's frame as the r-process calculations are done in this frame.  The results for the density evolution are then transformed to the observer's frame, including relativistic effects\footnote{To convert from the shell's frame to that of the observer for density and time the equations are:
\begin{equation}
\rho_{\rm obs} = \frac{\rho_{\rm shell}}{\Gamma}
\end{equation}

\begin{equation}
\delta t_{\rm obs} = \Gamma\,\delta t_{\rm shell}
\end{equation}

respectively.}. 
   
\begin{equation}
\label{shellexpansion}
 \rho = \frac{M_{\rm QN, ejecta}}{4 \pi} \left(R_{\rm NS} + \beta c \Gamma t  \right)^{-2} \Delta r^{-1} 
\end{equation}

 Compared to the simple pressure-driven expansion case as given in Eqn. \ref{radius}, the density drops much quicker, especially as ejecta speeds become relativistic. The corresponding expansion time-scale 
 $\tau_{\rm Rel.}\sim R_{\rm NS}/ c \, \Gamma$ is also much shorter (much less than 0.1 millisecond). 
Note that both the non-relativistic ($\beta={\rm v}/{\rm c}$, $\Gamma=1$) and relativistic ($\beta\simeq 1$, $\Gamma>1$) expansions are covered by Eqn. \ref{shellexpansion}. In the relativistic case the timestep $\delta t$ is adjusted accordingly to take into account the fast fall-off of the density.  The code switches automatically from the degenerate to non-degenerate expression by tracking the expression for the entropy/nucleon in time.

These calculations are embedded as an independent module into the code, so that both neutron star structure and r-process simulation are flexible, but addressed in a consistent manner.         

\subsection{Simulation Results}
\label{QNresults}

In \cite{jaik07}, the focus was on studying the dependence of the r-process yield on the following key parameters: $Y_e$ (crustal electron fraction), $\tau$ (expansion time-scale of the ejecta) and heating from nuclear reactions ($\beta$-decays). A full reaction network using the Clemson University nucleosynthesis code \citep{JM04} was coupled at the end stages of decompression to obtain the final abundance.  An important difference from \cite{jaik07} is that the expanding ejecta is now treated as being driven and heated by a shock wave from an underlying phase transition to quark matter \citep{LO}, thus the decompressing matter is not necessarily initially cold. Furthermore, the ejecta can be relativistic which was not considered in \cite{jaik07}. 

Following a similar methodology as when investigating the NSM scenario we chose to study the effect on r-process abundance yield of varying both the physical parameters as well as the initial nuclei abundance distributions.  The physical parameters of the QN investigated in this work are the mass of the ejecta and the percentage of QN energy transformed into kinetic energy of the ejecta.  The three initial abundance scenarios studied are; nuclei in ejecta are initially in NSE, the presence of an iron group seed and initial dissociation into neutrons, protons and alpha particles. 

\subsubsection{Varying Physical Parameters}
\label{QNphysRes}
As discussed in Sect. \ref{QNinput} the input parameters for an r-process simulation in the QN module are parameterized to be completely described by global NS properties (mass and radius), the mass of the ejecta ($m_{\rm ej}$) and the percentage of QN energy deposited as kinetic energy of the ejecta ($\zeta_{\rm KE}$).  For this analysis we chose to fix the mass and radius of the NS to canonical values (M$_{\rm NS} = 1.4$ M$_{\odot}$, R$_{\rm NS} = 10$ km).

\begin{figure}
\includegraphics[scale = 0.75]{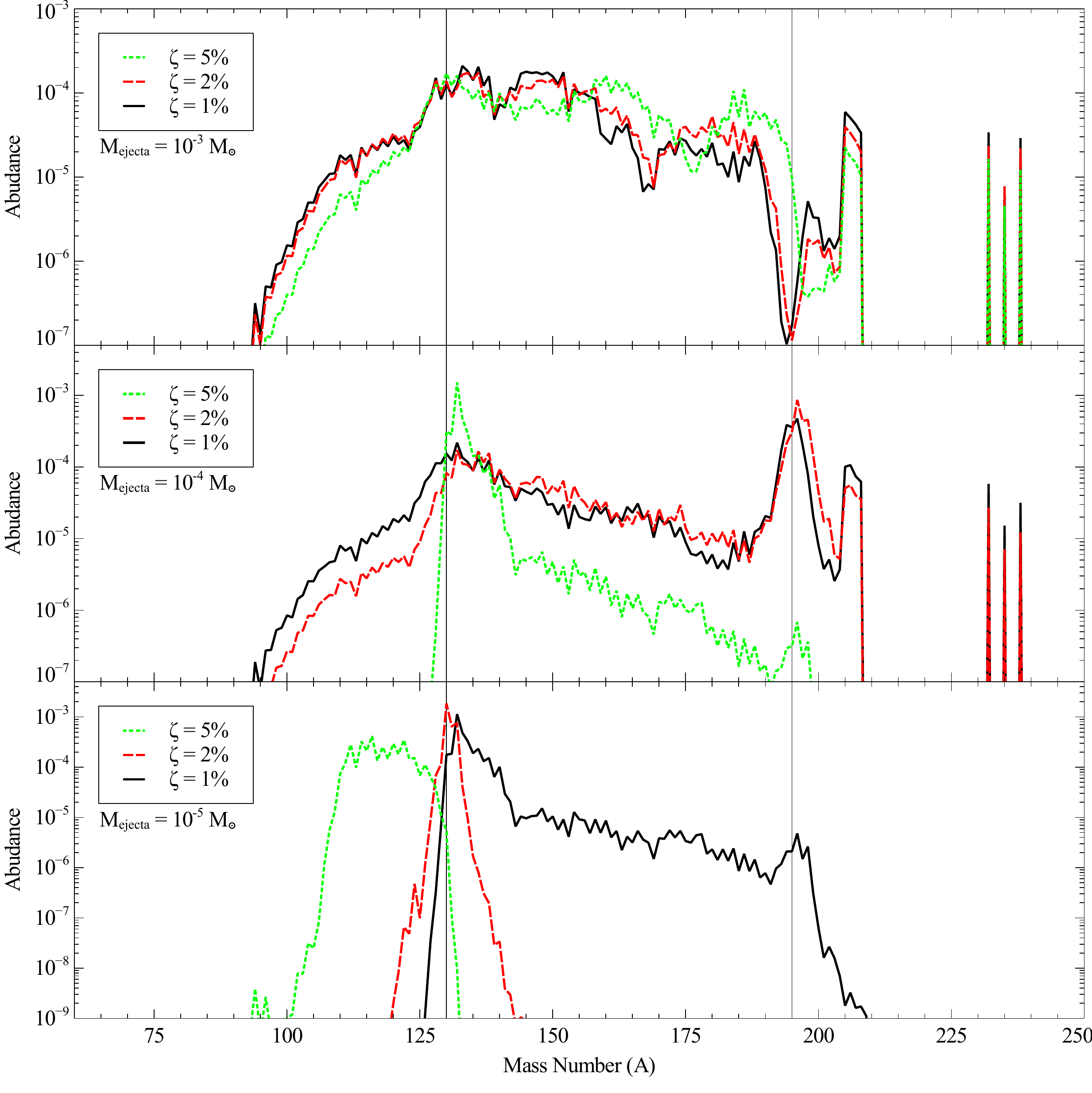}
\caption{\textbf{Quark nova.} Comparison of the effect on r-process yield of varying the percentage of quark nova energy transformed into kinetic energy of the ejecta.  In each panel the black solid line denotes 1$\%$, the red dashed line denotes 2$\%$ and the green dotted line 5$\%$.  For each panel the initial nuclei abundance distribution is determined by NSE.  \textit{Top:}  Quark nova ejecta of $10^{-3}$M$_{\odot}$.  \textit{Middle:}  Quark nova ejecta of $10^{-4}$M$_{\odot}$.
\textit{Bottom:}  Quark nova ejecta of $10^{-5}$M$_{\odot}$. }
\label{QNcompMejZeta}
\end{figure}

The effect of varying $\zeta_{\rm KE}$ can be seen in Fig. \ref{QNcompMejZeta}.  For this examination of the QN scenario we chose the initial abundance to be described by NSE.  For each panel of Fig. \ref{QNcompMejZeta} a different QN ejecta mass was chosen (top: $m_{\rm ej} = 10^{-3}$ M$_{\odot}$, middle: $m_{\rm ej} = 10^{-4} M_{\odot}$ and bottom: $m_{\rm ej} = 10^{-5} M_{\odot}$) and as described in Sect. \ref{QNinput} $m_{\rm ej}$ along with the global NS properties uniquely determine the input parameters for NSE ($T$, $Y_{\rm e}$ and $\rho$).

For the $m_{\rm ej} = 10^{-3} M_{\odot}$ case, seen in the top panel of Fig. \ref{QNcompMejZeta}, the effect of varying $\zeta_{\rm KE}$ is the smallest of all cases studied.  This being the largest ejecta mass chosen, from the description in Sect. \ref{QNinput}, this implies as well the slowest expansion.  With the slower expansion, for each $\zeta_{\rm KE}$ (1$\%$, 2 $\%$ and 5 $\%$) the neutron density remains sufficiently high for efficient capture.  The variance in final nuclei abundance is small comparing the results from the $\zeta_{\rm KE} = 1\%$ and $\zeta_{\rm KE} = 2\%$, with the $\zeta_{\rm KE} = 1\%$ case producing a slightly larger abundance of heavy (A $>$ 195) nuclei.  The production of heavier nuclei in the region of 130 $<$ A $<$ 195 in the $\zeta_{\rm KE} = 5\%$ simulation is due to the r-process not as readily reaching the fissionable region.  For the $\zeta_{\rm KE} = 1\%$ and $2\%$  cases, neutron-capture is capable of pushing through the A = 195 waiting point which implies more fission recycling occurs.  In the final nuclei abundance distribution this can be see in two ways; a higher abundance of nuclei in the region of A $<$ 130, which are all fission daughter nuclei in this case, and a larger abundance of A $>$ 195 nuclei.

For the $m_{\rm ej} = 10^{-4} M_{\odot}$ case, shown in the middle panel of Fig. \ref{QNcompMejZeta}, the final abundances of the $\zeta_{\rm KE} = 1\%$ and $2\%$ cases are quite similar, especially in the 130 $<$ A $<$ 195 region.  The main difference is that the slower expansion of the $\zeta_{\rm KE} = 1\%$ case compared to the $\zeta_{\rm KE} = 2\%$ simulation implies a higher maintained neutron density and thus the r-process is capable of proceeding to heavier nuclei in greater abundances.  This manifests in the final nuclei distribution as an increased production of $^{232}$Th, $^{235}$U and $^{238}$U as well as more fission daughter nuclei seen in the  $\zeta_{\rm KE} = 1\%$ results.  For the simulations results shown in the top and middle panel of Fig. \ref{QNcompMejZeta} all nuclei with A $<$ 130 in the final distribution are the result of fission.  With $\zeta_{\rm KE} = 5\%$ for the $m_{\rm ej} = 10^{-4} M_{\odot}$ case the rapid expansion of the ejecta causes neutron density to drop quickly.  The final nuclei distribution is dominated by nuclei slightly heavier than the A = 130 magic number with only low relative abundances of nuclei greater than A = 145.

For QN with ejecta of mass $m_{\rm ej} = 10^{-5} M_{\odot}$, changing $\zeta_{\rm KE}$ has the most significant impact on final nuclei abundance distribution of all cases studied in this work.  The low  mass implies that the ejecta will be travelling ultra-relativistically ($\Gamma_{\rm QN} \sim 300$ for $\zeta_{\rm KE} = 1\%$ and $\Gamma_{\rm QN} \sim 1500$ for $\zeta_{\rm KE} = 5\%$) and thus the density rapidly drops, adiabatically cooling the ejecta quickly.  The final abundances in the  $\zeta_{\rm KE} = 5\%$ case is changed only slightly from the initial NSE determined abundances.  With $\zeta_{\rm KE} = 2\%$ for $m_{\rm ej} = 10^{-5} M_{\odot}$ the r-process can proceed up to roughly A = 130, but the environment becomes unsuitable for r-process before the system is able to significantly push past that waiting point.  For $\zeta_{\rm KE} = 1\%$ the final nuclei abundance distribution contains a strong peak slightly to the heavy-side of A = 130 and small peak roughly centred at A = 195.  None of the $m_{\rm ej} = 10^{-5} M_{\odot}$ simulations were able to produce appreciable amounts of $^{232}$Th, $^{235}$U and $^{238}$U.

\subsubsection{Varying Initial Abundances}
\label{QNiniRes}
In our comparison of the effect of different initial abundances on r-process yield we chose to fix $\zeta_{\rm KE}$ to $1\%$ and once again set the global properties of the NS to canonical values.  For each initial nuclei distribution in the QN ejecta we compared three different ejecta masses $m_{\rm ej}\in\lbrace 10^{-5} M_{\odot}, 10^{-4} M_{\odot}, 10^{-3} M_{\odot} \rbrace$.

\begin{figure}
\includegraphics[scale = 0.75]{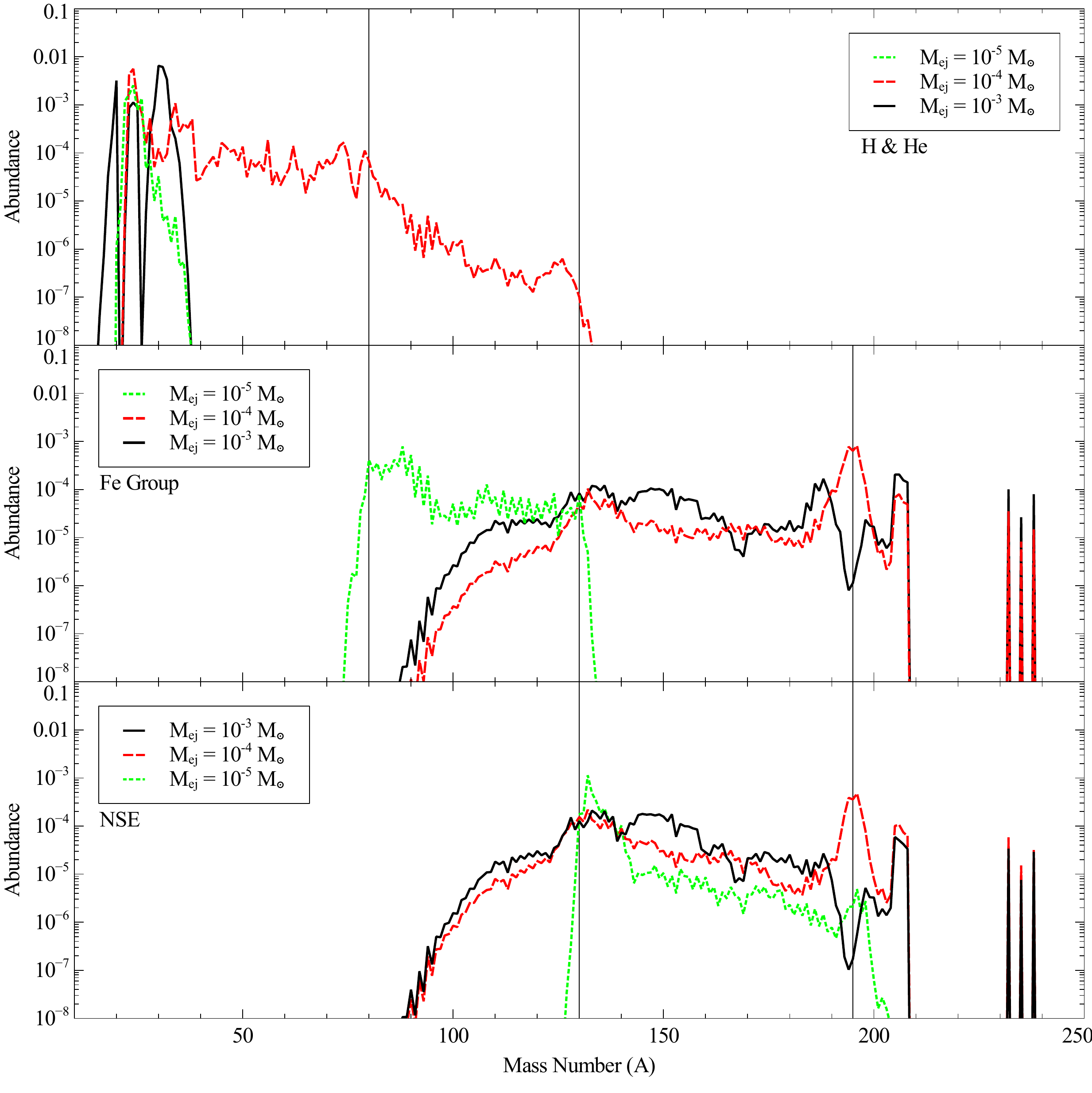}
\caption{\textbf{Quark nova.}  Comparison of the final r-process abundance yield of different masses of quark nova ejecta ($10^{-3}$M$_{\odot}$ denoted by the black solid line, $10^{-4}$M$_{\odot}$ by the red dashed line and $10^{-5}$M$_{\odot}$ by the green dotted line).  See section \ref{QNiniRes} for details of simulations.  \textit{Top:}  Initially only neutrons, protons and $\alpha$-particles.  \textit{Middle:} Initially a seed of iron group isotopes.  \textit{Bottom:}  Initial abundance determined by NSE.}
\label{QNcompInitial}
\end{figure}

A comparison of different $m_{\rm ej}$ in the QN scenario all starting with an initial abundance distribution consisting of neutrons, protons and alpha particles can be seen the top panel of Fig. \ref{QNcompInitial}.  The $m_{\rm ej} = 10^{-3} M_{\odot}$ implies the highest neutron-to-seed ratio of cases studied, however the r-process was only capable of creating nuclei up to A $\simeq 37$.  While the $m_{\rm ej} = 10^{-4} M_{\odot}$ simulation which started with a lower neutron-to-seed ratio was able to form nuclei up to A $\sim$ 130.   This result is due to the high temperature associated with the $m_{\rm ej} = 10^{-3} M_{\odot}$ ejecta causing photo-dissociation to inhibit the r-process, which is less of a factor for the $m_{\rm ej} = 10^{-4} M_{\odot}$ case.  The $10^{-5} M_{\odot}$ simulation produced an abundance peak at A $\sim 24$ with a heavy-side tail, all of which is the result of neutron capture onto $^{20}$Ne.  As stated in Sect. \ref{NSMiniRes}, this work should be revisited with a more robust treatment of alpha-capture.

The final nuclei abundances for simulations of the set $m_{\rm ej} \in \lbrace 10^{-3} M_{\odot}, 10^{-4} M_{\odot}, 10^{-5} M_{\odot} \rbrace$ starting from a seed of iron group isotopes are plotted in the middle panel of Fig. \ref{QNcompInitial}.  The rapid expansion of the  $m_{\rm ej} = 10^{-5} M_{\odot}$ ejecta does not allow for the r-process to proceed past the A = 130 waiting point and a strong peak shifted to the heavy-side of A = 80 waiting point dominates the final abundance distribution in the $m_{\rm ej} = 10^{-5} M_{\odot}$ case.  For both the $m_{\rm ej} = 10^{-4} M_{\odot}$ and $10^{-3} M_{\odot}$ simulations a significant amount of fission recycling occurs which shapes the final nuclei abundance distribution.  The final abundance of the $m_{\rm ej} = 10^{-4} M_{\odot}$ case displays a strong peak at A = 195 and lesser, broader peak slightly shifted to the heavy-side of the A = 130 solar peak.  The higher neutron-to-seed ratio for $m_{\rm ej} =  10^{-3} M_{\odot}$ allows for a greater degree of fission recycling when compared to the other two cases shown in the middle panel of Fig. \ref{QNcompInitial}, which manifest in the final nuclei distribution as an increase abundance of nuclei below A = 130 as well as the production of the largest abundances of $^{232}$Th, $^{235}$U and $^{238}$U.

The bottom panel of Fig. \ref{QNcompInitial} displays the final abundances for the aforementioned set of $m_{\rm ej}$, all of which started from an initial abundance distribution described by NSE.  While each of the plotted abundances in the bottom panel of Fig. \ref{QNcompInitial} appear separately in Fig. \ref{QNcompMejZeta} comparing them directly illuminates the trend that increasing $m_{\rm ej}$ implying lowering of Y$_{\rm e,}$ allows for the r-process to proceed to heavier nuclei. The final nuclei distribution of the $m_{\rm ej} = 10^{-4} M_{\odot}$ contains the most heavy nuclei (A $>$ 195).  However the increased abundance of nuclei with A $\leq$ 130 (which are all fission daughter nuclei in these cases) in the $m_{\rm ej} = 10^{-3} M_{\odot}$ is due to that fact that a greater amount of super-heavy (A $>$ 250) nuclei where reached during the r-process that then subsequently fissioned during the decay to stability. 

\section{Comparison to Solar Abundances}
\label{discuss}
In order to compare the r-process abundances from each of our studied astrophysical scenarios to observed solar abundances we consider a superposition of computed results within each scenario.  The results chosen for this comparison do not represent an exhaustive search of the available parameter-space for each astrophysical site.  Instead for this comparison to solar we chose simulation results from our parameter survey of each site that best reproduced the observed solar distribution.  We leave an exhaustive parameter survey to future users of r-Java 2.0.  

\begin{figure}
\includegraphics[scale = 0.85]{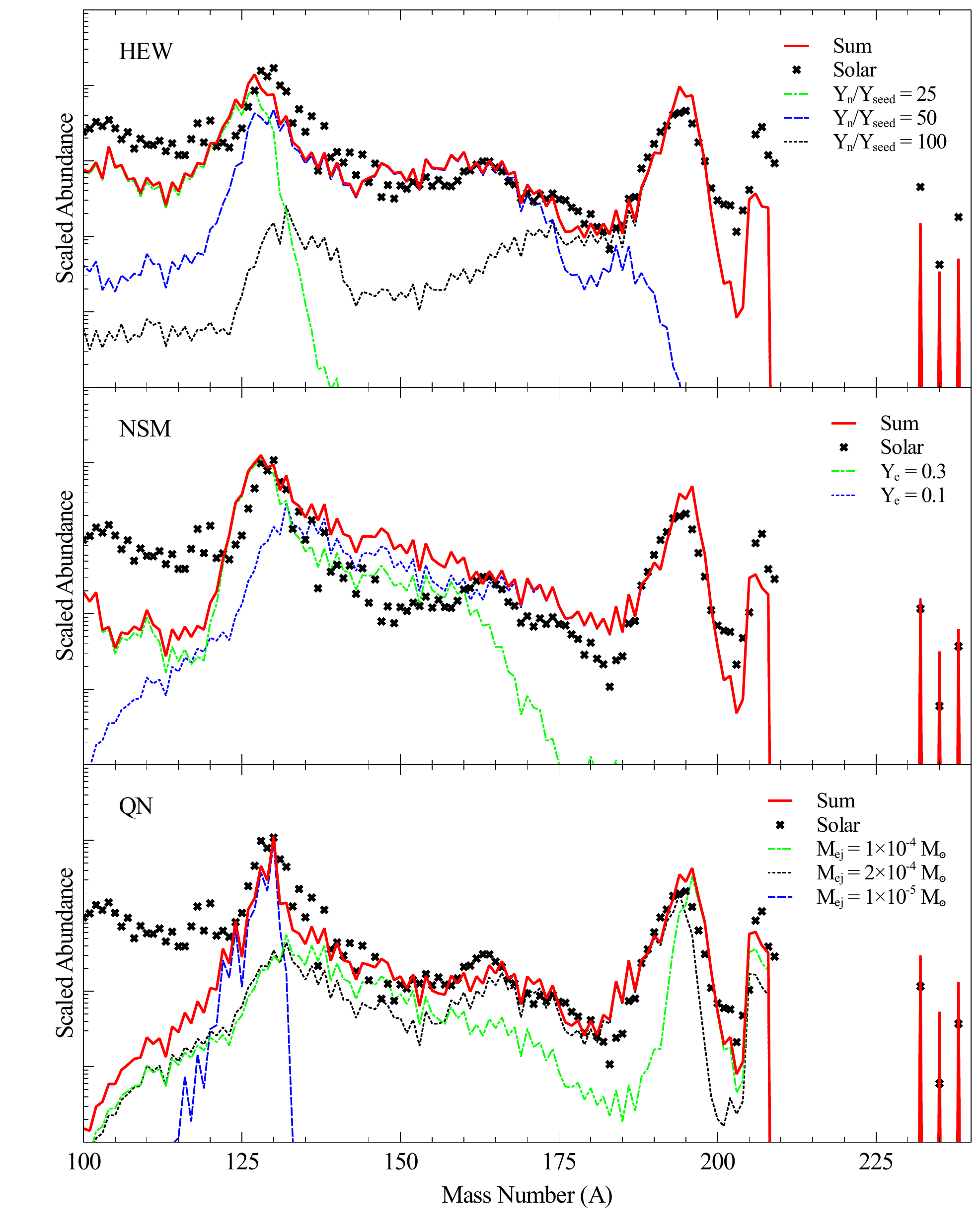}
\caption{In each panel the superposition of the r-process yields (red solid line) from each astrophysical site is compared to the observed solar r-process abundances (black crosses).  The constituent components are as well plotted for each site.  \textit{Top:} The r-process abundances from high-entropy wind events; $Y_{\rm n} / Y_{\rm seed} = 25$ (green dash-dot line), $Y_{\rm n} / Y_{\rm seed} = 50$ (blue dashed line) and  $Y_{\rm n} / Y_{\rm seed} = 100$ (black dotted line). \textit{Middle:}  The r-process abundances from neutron star merger events; $Y_{\rm e}  = 0.3$ and $\rho = = 3 \times 10 ^{11}$g cm $^{-3}$ (green dash-dot line), $Y_{\rm e}  = 0.1$ and $\rho = 8 \times 10 ^{11}$g cm $^{-3}$ (blue dashed line).  \textit{Bottom:}   The r-process abundances from quark novae; $m_{\rm ej} = 10^{-4} M_{\odot}$ and $\zeta = 1\%$ (green dash-dot line), $m_{\rm ej} = 10^{-4} M_{\odot}$ and $\zeta = 5\%$ (black dotted line) and $m_{\rm ej} = 10^{-5} M_{\odot}$ and $\zeta = 1\%$ (blue dashed line).  }
\label{sol}
\end{figure}

The superposition from a set of HEW r-process abundances from different initial conditions is compared to the solar abundances in the top panel of Fig. \ref{sol}.  Each of the contributing HEW scenario began from an $\alpha$-rich freeze-out initial composition, however each consider different neutron-to-seed ratio (25, 50, 100).  When compared to the solar abundances, the superposition of these HEW abundances overproduces the A = 195 peak slightly while under-producing the heavy-side of the A = 130 peak.  The basic shape of the 145 $<$ A $<$ 180 region of the solar abundance, which slowly rises and peaks a A$\sim165$ and drops more sharply until A$\sim182$ is captured in the HEW scenario.  

A comparison to the solar r-process abundance of a superposition of two simulation runs in the NSM scenario, one with $Y_{\rm e,0}$ = 0.3 and the other $Y_{\rm e,0}$ = 0.3, both starting from $T_9 = 3$ and $\rho = 8 \times 10^{11}$g cm$^{-3}$ with a initial nuclei distribution calculated using NSE can be seen in the middle panel of Fig. \ref{sol}.  The breadth of the solar A = 130 peak is well reproduced by the NSM abundance.  The shape of the A = 195 solar peak is as well captured by the NSM r-process results, although the height of the A = 195 relative to the A = 130 peak is higher than that observed in the solar abundances.  The region between the A = 130 and A = 195 peaks is overproduced by the superposition of NSM abundances.  

The bottom panel of Fig. \ref{sol} displays a comparison of the superposition of the r-process yield from three QN ejecta scenarios ($M_{\rm QN, ejecta} \in \lbrace1\times10^{-4} M_{\odot},2\times10^{-4} M_{\odot},1\times10^{-5} M_{\odot}\rbrace$) to the observed solar r-process abundances.  All of the QN scenarios begin with the nuclei distribution computed using NSE and considered $\zeta = 1\%$.  The superposition of the QN r-process abundances under-produce the breadth of the solar A = 130 peak, but reproduces the europium rare earth peak seen in the solar abundances.  The breadth of the solar A = 195 peak is achieved by the QN scenario, although the relative height compared to the A = 130 peak is high compared to that seen in the solar distribution. 

The europium rare earth peak observed in solar abundances is reproduced in both the HEW and QN scenarios, but not for NSM.  This difference is due to the fact that for the NSM scenario the density and temperature have dropped much lower prior to neutron freeze-out compared to the other two scenarios.  For this reason in the NSM scenario beta-decay compete with neutron-capture prior to freeze-out which acts to smooth out the rare earth peak \citep{surman97}.  Recent study into fission fragmentation of neutron-rich isotopes by \cite{gori13} have found a new possible origin for the europium rare earth peak that is not considered in this work.
 
\begin{figure}
\includegraphics{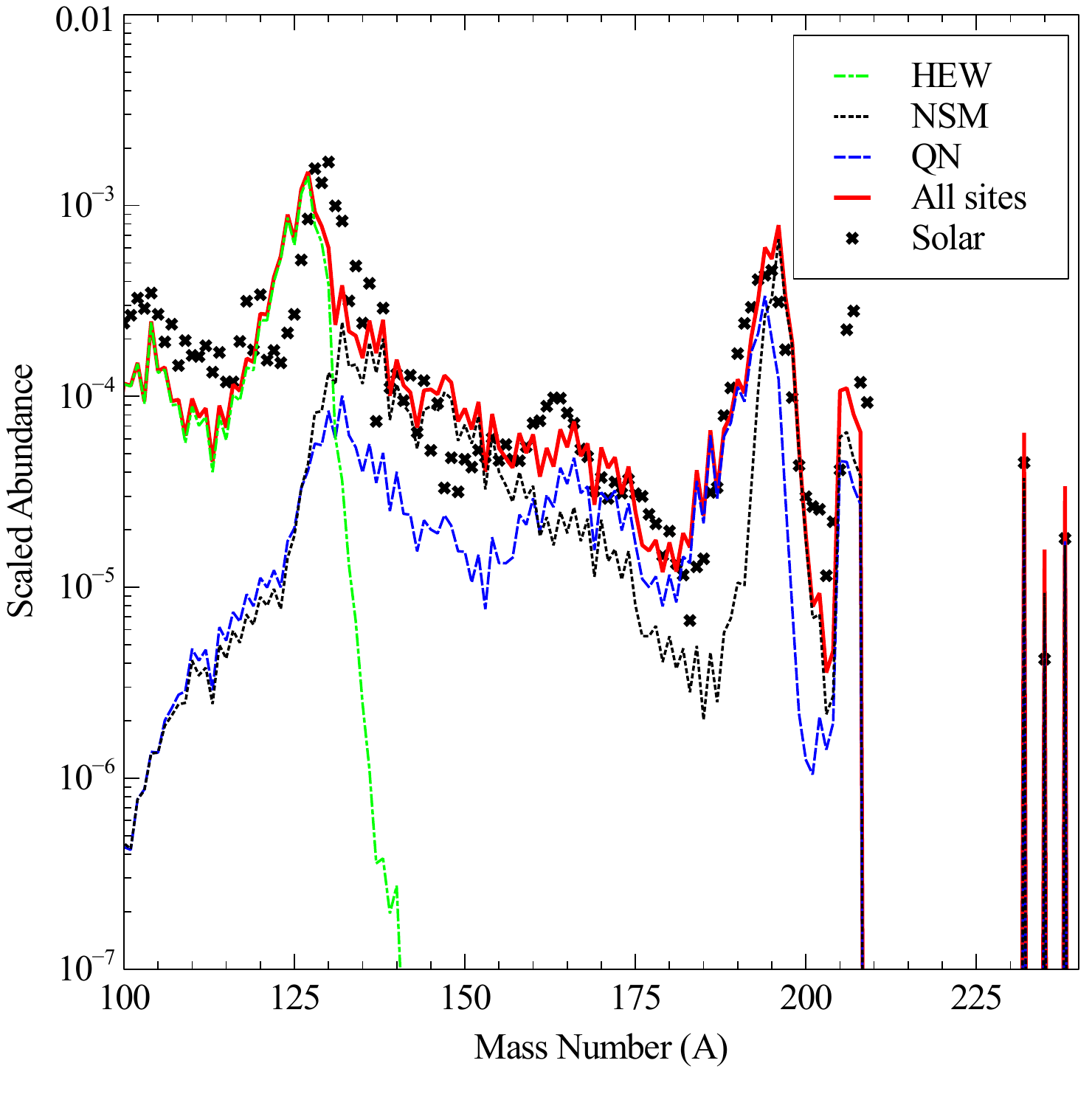}
\caption{All sites scaled by predicted rates compared to solar abundances. }
\label{allSites}
\end{figure}

Displayed in Fig. \ref{allSites} is a consideration of the contribution from each of the astrophysical sites studied in this work in comparison with observed solar r-process abundances.   A representative member from each site was chosen such that the scaled superposition best resembled the solar r-process abundance.  From the HEW scenario the results from the $\alpha$-rich freeze-out simulation with a neutron-to-seed ratio of 25 was used.  The QN results used in Fig. \ref{allSites} considered the ejection of $2\times10^{-4} M_{\odot}$ from a 1.8 $M_{\odot}$ neutron star.  The final r-process abundance from the ejecta initially at $T_9 = 6$ and $Y_{\rm e}$ = 0.1 was used as the contribution from the NSM in Fig. \ref{allSites}.  Each of the the contributing sites were scaled by their local predicted rates before considering the superposition.   A cosmic core-collapse supernova rate of $7\times10^{-5}$ yr$^{-1}$ within 1 Mpc predicted from the cosmic star formation rate from \cite[][and references therein]{snr} was used with the HEW abundance.  The abundance from the NSM was multiplied by the expected NSM rate within our Galaxy of $10^{-5}$yr$^{-1}$ \citep{phin91,arz99,blec02}.  For the QN contribution, the prediction that $\sim10\%$ \citep{ODD} of core-collapse supernovae may be followed by a QN was used.

\section{Summary and Conclusions}
\label{conc}
For the HEW simulations done in this work when the initial nuclei abundances were calculated under the assumption of NSE the r-process was not capable of producing nuclei significantly heavier that A = 80.  When an $\alpha$-rich freeze-out was assumed to compute the initial abundances a strong A = 130 peak was synthesized for all parameters surveyed.  Starting from an $\alpha$-rich freeze-out only the largest entropy (S = 270 and thus neutron-to-seed ratio = 100) studied was capable of producing transuranic elements.  The HEW scenario reproduced the europium rare earth peak observed in solar r-process abundances.

Fission recycling had the most impact on final r-process abundances in the NSM of all the scenarios studied.  For all parameters surveyed (with the exception of the initial abundance of only neutrons, protons and alpha particles) r-process in the context NSM was capable of synthesizing the A = 195 peak and in most cases transuranic elements as well.  When compared to the solar abundances NSM robustly reproduced the A = 130 and A = 195 peaks, however the europium rare earth peak was not formed.  Recent work by \cite{gori13} studying fission fragmentation of neutron-rich nuclei has addressed the europium rare earth peak in the context of NSM.

Simulations done in this work showed that an ejecta mass of $10^{-4} M_{\odot}$ produced the strongest A =195 of all QN scenarios studied.  The lighter ejecta mass ($10^{-5} M_{\odot}$) studied cooled too rapidly for the r-process to be able to produce transuranic elements.  The final r-process abundances in heavier ($10^{-3} M_{\odot}$) ejecta case produced a relatively flat distribution of nuclei.  When compared to solar abundances, the QN scenario under-produced the breadth of the A = 130 peak but was capable of synthesizing the rare earth peak and also the A = 195 peak.   

For the simulations performed in this work no single parameter set for a chosen astrophysical site could reproduce the solar r-process abundance distribution.  Either a combination of different astrophysical sites or different parameter sets within an astrophysical site was needed in order to re-produce solar abundances.  Within the parameters surveyed in this work each scenario is capable of producing both a A = 130 and A = 195 peak comparable to those observed in the solar abundances.  

For the parameters studied in this work, the resultant r-process abundances for the NSM and QN were quite similar.  This similarity is due to the high initial neutron-to-seed ratio for both scenarios which causes fission recycling to become a driving factor that shapes the final nuclei distribution.  The similarity in r-process yields from these two scenarios could be an indication as to a cause for the apparent universality of the r-process nuclei distribution \citep{qianUni}.

The study of r-process nucleosynthesis remains a challenging topic which requires a deeper understanding of both the underlying nuclear physics as well as the astrophysics that shapes the r-process environment.  With r-Java 2.0 we have provided a robust and easy-to-use platform that is capable of tackling both sides of the r-process problem.  In \cite{kostka} we discussed the cutting-edge nuclear physics incorporated in r-Java 2.0 which includes; temperature-dependant neutron capture cross-sections and photo-dissociation rates, realistic fission recycling and beta-delayed neutron emission.  In this work we have shown the capabilities of r-Java 2.0 to model the evolution of density and temperature of three different proposed astrophysical r-process site; HEWs around a proto-neturon star, ejecta from NSMs, and ejecta from QNe. 

A fundamental tenet followed during the development of r-Java 2.0 was maximizing flexibility.  For this reason we ensured that the software is cross-platform compatible and gave the user the ability to change any nuclear property both quickly and easily.  In this work we have used the three built-in astrophysical modules in order to highlight the fact that with r-Java 2.0 for the first time anyone can run a comparison of proposed astrophysical r-process sites with their dynamical evolution and ejecta conditions.  R-Java 2.0 goes beyond existing codes by including the option to define a custom density evolution profile in order to study other possible r-process scenarios.

As many of the nuclear inputs used in r-Java 2.0 are only theoretically known, the results from rare-isotope beam facilities and upcoming sensitivity studies will provide much needed experimental insight into  r-process nuclei \citep{dillman,hosmer,vanscelt}.  The flexibility of r-Java 2.0 allows for researchers to easily include experimental results in order to test their impact on the r-process.  

The areas of development that we are undertaking for the next release of r-Java are;  the inclusion of neutrino-induced and spallation reactions, the development of a charged-particle reaction network module and adding the ability to study nuclear isomers.

\onecolumn

\clearpage




\clearpage

\clearpage




\clearpage




\clearpage

\clearpage




\clearpage



\clearpage

\clearpage
\begin{table}
\caption{Quark Nova Ejecta Parameters} 
\label{qnTable} 
\centering 
\begin{tabular}{c | c c c c} 
\hline\hline 
Mass & T$_{0}$ & $\rho_0$ & Y$_{\rm e,0}$ & $\tau$ \\ 

[M$_{\odot}$]& [10$^9$K] &  [10$^{11}$g cm$^{-3}$] & &[$\mu$s] \\
\hline 
10$^{-5}$ & 2.4 & 1.0 & 0.18 & 0.1 \\ 
10$^{-4}$ & 3.8 & 4.0 & 0.13 & 0.5 \\ 
10$^{-3}$ & 7.5 & 30 & 0.07 & 6.0 \\ 
\hline 
\end{tabular}
\end{table}

\clearpage
\appendix
\section{Derivation of QN shell thickness evolution}
In the frame of the QN ejecta shell, the thickness ($\Delta r$) evolves linearly with the sound speed ($c_{\rm s} = \kappa \rho^{\nu}$).  For relativistically degenerate matter ($\rho > 2 \times 10 ^ 6$ g cm$^{-3}$) $\kappa = 2.57 \times 10^7$ and  $\nu = 1/6$ and for densities below this transition point $\kappa = 2.26 \times 10^6$ and  $\nu = 1/3$,

\begin{equation}
\label{initThick}
d\left(\Delta r\right) = \kappa \rho^{\nu} dt \, .
\end{equation}

\noindent Rewriting $\rho$ in terms of ejecta mass ($m_{\rm ej}$) Eqn. \ref{initThick} becomes,

\begin{equation}
\label{2Thick}
\begin{aligned}
d\left(\Delta r\right) = \kappa \,dt \, \left( \frac{4 \pi m_{\rm ej}}{r^2 \Delta r} \right)^{\nu} \left( \frac{\Delta r_0}{\Delta r_0} \right)^{\nu} \left( \frac{R_{\rm NS}}{R_{\rm NS}} \right)^{2\nu} \\
d\left(\Delta r\right) =\kappa \, dt \,\rho _{0}^{\nu} \left( \frac{\Delta r_0}{\Delta r} \right)^{\nu} \left( \frac{R_{\rm NS}}{r} \right)^{2\nu}
\end{aligned}
\end{equation}

\noindent Rearranging Eqn. \ref{2Thick} and expressing the evolution of shell radius as $r = R_{\rm NS} + \beta c \Gamma t$ gives;

\begin{equation}
\label{3Thick}
\left( \frac{\Delta r}{\Delta r_0} \right)^{\nu} d\left( \Delta r \right) = \kappa \, \rho_0^{\nu} \frac{dt}{\left( 1 + t/\tau \right)^{2\nu}}\, ,
\end{equation} 

\noindent where $\tau = R_{\rm NS} / \left( \beta c \Gamma \right)$ defines the expansion timescale.  Integrating Eqn. \ref{3Thick} yields,

\begin{equation}
\label{4Thick}
\left( \frac{\Delta r}{\Delta r_0} \right)^{\nu + 1} =  \left( \frac{\nu + 1}{\Delta r_0}\, \right) \kappa \, \rho_0^{\nu} \, \left( \frac{\tau}{1 - 2 \nu} \right) \left( 1 + \frac{t}{ \tau} \right)^{1-2\nu} + C \, ,
\end{equation}

\noindent where the constant of integration ($C$) can be found from initial conditions ($\Delta r(t=0) = \Delta r_0$).  This gives an expression for the evolution of shell thickness which is dependant on the degeneracy of the matter;

\begin{equation}
\label{5Thick}
\Delta r = \left(\Delta r_0 ^{\nu +1} + \left( \frac{\nu + 1}{1 - 2\nu} \right)\, \kappa \, \rho_0^{\nu}\, \Delta r_0^{\nu}\, \tau \, \left( \left( 1 + \frac{t}{\tau} \right)^{1-2\nu} - 1  \right)  \right)^{1/(\nu + 1)} \, .
\end{equation}

\noindent The expression for shell thickness evolution is used in Eqn. \ref{shellexpansion} to determine the density evolution of the QN ejecta. 

\clearpage
\bibliographystyle{aa}
\bibliography{hew}

\end{document}